\documentclass[conference]{IEEEtran}
\IEEEoverridecommandlockouts
\usepackage{cite}
\usepackage{amsmath,amssymb,amsfonts}
\usepackage{algorithmic}
\usepackage{graphicx}
\usepackage{textcomp}
\usepackage{booktabs} 
\usepackage{xcolor}
\usepackage{caption}
\usepackage[margin=1in]{geometry}
\usepackage{graphicx}
\usepackage{url}
\usepackage{amsmath}

\usepackage{graphicx} % for \resizebox

\usepackage{booktabs}
\usepackage{tabularx}
\usepackage{caption}
\usepackage{multirow} 
\usepackage{array}  % In preamble
\usepackage{setspace}
\usepackage{graphicx}
\usepackage{wrapfig}
\usepackage{graphicx}
\def\BibTeX{{\rm B\kern-.05em{\sc i\kern-.025em b}\kern-.08em
    T\kern-.1667em\lower.7ex\hbox{E}\kern-.125emX}}
\begin{document}

\title{The GenAI Generation: Student Views of Awareness, Preparedness, and Concern\\
}

% \author{\IEEEauthorblockN{1\textsuperscript{st} REDACTED}
% \IEEEauthorblockA{\textit{REDACTED} \\
% REDACTED\\
% REDACTED}
% \and
% \IEEEauthorblockN{2\textsuperscript{nd} REDACTED}
% \IEEEauthorblockA{\textit{REDACTED} \\
% REDACTED\\
% REDACTED}
% \and
% \IEEEauthorblockN{3\textsuperscript{rd} REDACTED}
% \IEEEauthorblockA{\textit{REDACTED} \\
% REDACTED\\
% REDACTED}
% }

\author{\IEEEauthorblockN{1\textsuperscript{st} Micaela Siraj}
\IEEEauthorblockA{\textit{Georgia Institute of Technology} \\
Atlanta, United States of America \\
msiraj3@gatech.edu}
\and
\IEEEauthorblockN{2\textsuperscript{nd} Jon Duke}
\IEEEauthorblockA{\textit{Georgia Institute of Technology} \\
Atlanta, United States of America\\
Jon.Duke@gatech.edu}
\and
\IEEEauthorblockN{3\textsuperscript{rd} Thomas Plötz}
\IEEEauthorblockA{\textit{Georgia Institute of Technology} \\
Atlanta, United States of America\\
Thomas.Ploetz@gatech.edu}
}

\maketitle

\begin{abstract}
Generative Artificial Intelligence (Gen AI) is revolutionizing education and workforce development, profoundly shaping how students learn, engage, and prepare for their future. Outpacing the development of uniform policies and structures, Gen AI has heralded a unique era and given rise to the Gen AI Generation, a cohort of individuals whose development has been increasingly shaped by the opportunities and challenges Gen AI presents during its widespread adoption within society. This study examines higher education students' perceptions of Gen AI through a concise survey with optional open-ended questions, focusing on their awareness, preparedness, and concerns. "Readiness" appears increasingly tied to exposure to Gen AI through one's coursework. Students with greater curricular exposure to Gen AI tend to feel more prepared, while those without more often express vulnerability and uncertainty, highlighting a new and growing divide that goes beyond traditional disciplinary boundaries. Evaluation of more than 250 responses, with over 40 percent providing detailed qualitative feedback, reveals a core dual sentiment: while most students express enthusiasm for Gen AI, an even greater proportion voice a spectrum of concerns about ethics, job displacement, and the adequacy of educational structures given the highly transformative technology. These findings offer critical insights into how students view the potential and pitfalls of Gen AI for future career impacts.
\end{abstract}
\section{Introduction}
The Gen AI Revolution represents a paradigm shift in technology adoption, enhancing efficiency, creativity, and performance while introducing challenges such as ethical concerns and the need for responsible stewardship~\cite{delios2024GenAI,bastani2024learning,kumar2025GenAI,basiouny2024guardrails}. This study introduces \textit{the Gen AI Generation}. The Gen AI Generation is defined as the cohort of individuals actively undergoing developmental and educational growth/attainment during the experimental, unregulated rise of Gen AI. Unlike earlier generations who did not have these tools and later generations who may inherit more structured systems, this group faces unique challenges and opportunities surrounding uncertainty.
\newline\textit{Developmental Context:} Actively undergoing critical cognitive and social growth while navigating the Gen AI Revolution. \textit{Navigating Uncertainty:} Early adopters, engaging with experimental, unregulated technologies without widely established policies. \textit{High Adaptability:} Balancing both the traditional and new competencies required to effectively interact and engage with Gen AI in a highly dynamic environment. \textit{Long-Term Impact Unknown:} Using these tools during formative years without a clear understanding of potential long-term cognitive or social consequences.
\newline Just as Google transformed how students accessed and retained information, dubbed the ``Google Effect''~\cite{sparrow2011google}, and the rise of digital technologies sparked concerns about ``digital dementia''~\cite{spitzer2012digital}, the Gen AI Revolution represents another critical shift. Evaluating the Gen AI Generation is essential to understand how these tools influence development, ensuring we can build a foundation of guidance and informed policies.
\newline
\textbf{Awareness of Gen AI’s Capabilities and Applications}
\newline 
For this study, students’ \textit{awareness} of GenAI refers to their familiarity with Gen AI tools and their understanding of how such tools can support academic and professional work. Prior research shows high global awareness and frequent use for tasks such as information retrieval and paraphrasing, though concerns about academic integrity and creativity remain~\cite{yusuf2024GenAI}. Engagement with GenAI also varies across demographic groups, indicating a need for targeted educational support~\cite{daher2024perceptions}. While students value GenAI for idea generation and brainstorming, many remain hesitant to use it for completing assignments, underscoring the need for clear ethical guidance~\cite{barrett2023ethics}.
\newline
Work on GenAI’s educational potential highlights its promise for personalized learning and adaptive tutoring that foster higher-order cognitive skills~\cite{chen2022tutoring}. Additional studies note that while personalized support tools can aid learning, over-reliance may hinder critical thinking, reinforcing the need for balanced instructional approaches~\cite{kutty2024criticalthinking}. Current research shows student awareness is shaped by geo-demographic factors, suggesting that diverse and ethically grounded strategies are needed to prepare students for GenAI’s role in education.
\newline
\textbf{Preparedness for an Gen AI-Driven Workforce}
\newline 
For this study, student \textit{preparedness} refers to their confidence in navigating an AI-driven job market and their sense of how well their education equips them for that future. Research identifies persistent gaps in AI-focused training within higher education, leaving many students feeling under-prepared, especially amid ethical and accuracy concerns that hinder wider integration~\cite{abdelwahab2022training, chan2023integration}. Scholars recommend embedding AI literacy, practical applications, and interdisciplinary, project-based learning to better align academic preparation with workforce demands~\cite{chiu2024learning}.
\newline
Faculty readiness is also critical: professional development programs can help ensure GenAI adoption supports institutional goals and strengthens critical digital literacy for both educators and students~\cite{newell2024facultydev}. Personalized learning supported by GenAI may further enhance academic and employability outcomes, though concerns about ethics and impacts on critical thinking remain~\cite{egeli2024personalized}. Overall, student preparedness underscores the need for intentional, interdisciplinary engagement with GenAI in the classroom, alongside adequate support for educators integrating these tools.
\newline
\textbf{Concerns About Gen AI’s Potential Implications}
\newline 
For this study, students’ \textit{concerns} refer to their anxiety or worry about GenAI’s potential impact. Research highlights ethical risks, such as data privacy, bias, and misinformation, alongside worries about overreliance and uncritical acceptance of AI outputs~\cite{barrett2023ethics, aksoy2024ethics}. Students express particular reservations using GenAI for assignments without disclosure, and studies warn that misuse may undermine academic integrity and originality, underscoring the need for critical evaluation skills and clear institutional guidelines~\cite{yusuf2024GenAI, aksoy2024ethics}~\cite{luo2024assessment}. Related work warns that over-reliance on GenAI could weaken critical thinking and foundational learning skills, calling for balanced and intentional integration into educational practices~\cite{kutty2024criticalthinking}. While GenAI can support advanced cognitive processes, persistent concerns about ethics, bias, and cognitive impacts highlight the need for coherent guidance that promotes responsible use while safeguarding student development~\cite{barrett2023ethics, yusuf2024GenAI, aksoy2024ethics, luo2024assessment, kutty2024criticalthinking}.
\newline \textbf{Contributions of This Study}
\newline 
This study provides insight into how higher education students are experiencing the rapid rise of Gen AI, introducing the term “Gen AI Generation”. Drawing on structured survey data and qualitative responses, the findings reveal patterns in students’ awareness, preparedness, and concern, showing a clear desire for intentional support. 
\section{Methods}
The survey included eight main questions, three demographic items, and two open-ended prompts. The study received Georgia Tech (GT) IRB exemption, and all participants provided informed consent. Designed for completion in under five minutes~\cite{galesic2009questionnaire}, the survey opened with required disclosures and a screening item, progressing from general to more sensitive topics~\cite{perreault1975order}. Items were refined through lab feedback and internal validation. Attitudes were measured using a four-point forced-choice Likert-type scale, intentionally omitting a neutral option to encourage engagement and reduce satisficing~\cite{joshi2015likert, krosnick2018attitudes}. Survey administration and full question wording are provided in the supplemental materials.
\newline
Two GT cohorts were sampled. The Fall 2023 Computer Science Graduate cohort included online ($\approx$12,000 students\cite{college2024omscs}) and on-campus students ($\approx$14,500\cite{collegeFacts}), recruited via listservs from Sept. 28–Nov. 1, 2023. The Fall 2023 Undergraduate cohort was recruited through the Office of Academic Effectiveness using stratified random sampling of 5,655 undergraduates balanced by gender, race, citizenship, and class standing (2\% margin of error). The survey ran Nov. 2–23, 2023 with reminder emails on Nov. 2, 15, and 22. Demographic breakdowns appear in the supplemental materials. 
Analysis included descriptive statistics and frequency distributions to compare subgroup patterns, supplemented by qualitative analysis of open-ended responses. For qualitative data, preprocessing involved tokenization, lowercasing, removal of punctuation and stop words, and extraction of unigrams through quadgrams with low-frequency terms excluded. Grammatical parsing of noun and verb phrases supported inductive coding to identify recurring themes.
\section{Results}
The undergraduate cohort included 115 respondents across three categories: 28 in Computer Science, 60 in Engineering, and 27 in Other disciplines. The graduate cohort response count was 102 and comprised solely of computer science majors. The full breakdown between sample to response counts can be seen in the supplemental materials.
\begin{figure}[tbp!] 
    \centering
    \setlength{\belowcaptionskip}{-6pt} \includegraphics[width=0.48\textwidth]{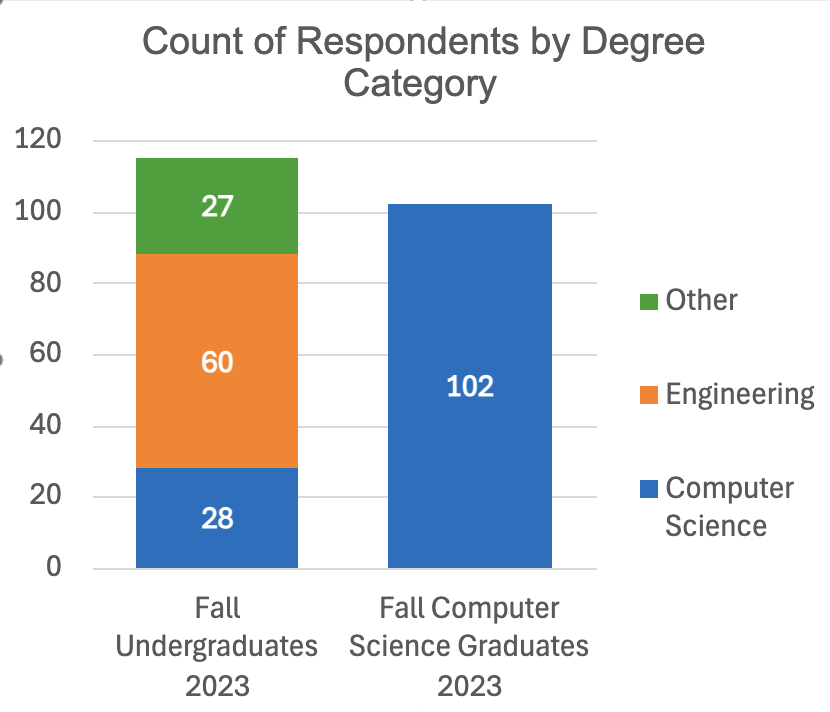}  % Replace with your figure file
    \vspace{-3pt}\caption{Response counts by degree and level}
    \label{fig:degree_count}
\end{figure}
\begin{figure}[tbp!]
    \centering
    \setlength{\belowcaptionskip}{-10pt}   \includegraphics[width=0.48\textwidth]{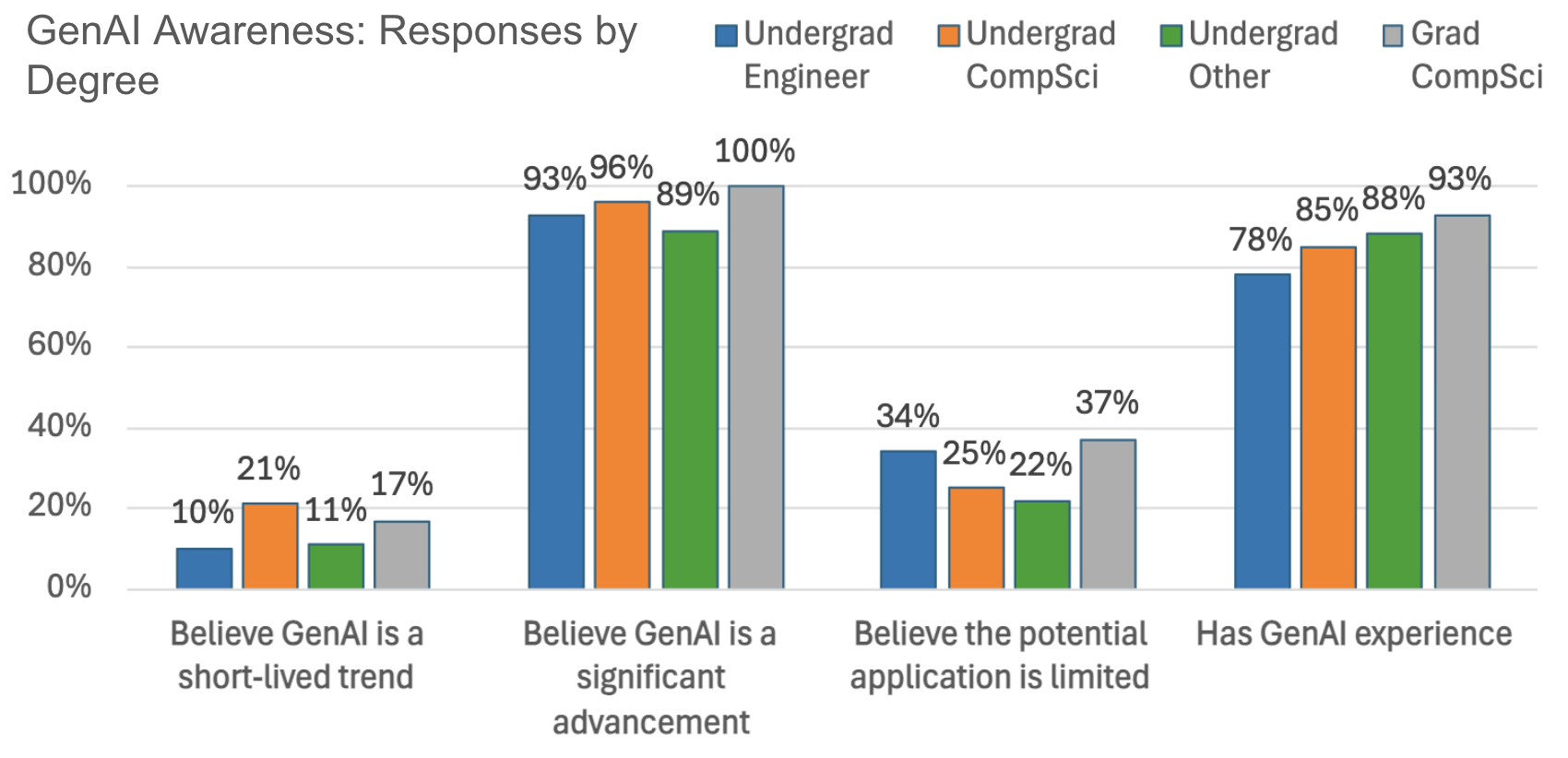}  % Replace with your figure file
    \vspace{-3pt}
    \caption{Responses on Gen AI awareness.}
    \vspace{-5pt}
    \label{fig:GenAI-awareness}
\end{figure}
\newline\textbf{Awareness}:  A small percentage of respondents in all groups believe Gen AI is a short-lived trend, seen in Figure 2, with undergraduate computer science students slightly more likely to agree at 21\%. Most respondents view Gen AI as a significant advancement, with nearly unanimous agreement among the graduate cohort at 100\%, followed by undergraduate computer science and engineers at 96\% and 93\%, respectively. When asked about the limitations of Gen AI's potential applications, agreement levels vary, with engineers and graduate respondents showing slightly higher agreement at 34\% and 37\%. Gen AI experience is highest among Graduate respondents (93\%) and lowest among Engineers (78\%).
\newline
\textbf{Preparedness}: For the percentage of respondents that believe job skills will change due to Gen AI, Figure 3 shows undergraduate computer science students have the highest percentage at 82\%, followed by computer science graduates at 77\%. Other majors report 56\%, while engineering students have the lowest percentage at 38\%.
\begin{figure}[tbp!]
    \centering
    \setlength{\belowcaptionskip}{-10pt} \includegraphics[width=0.48\textwidth]{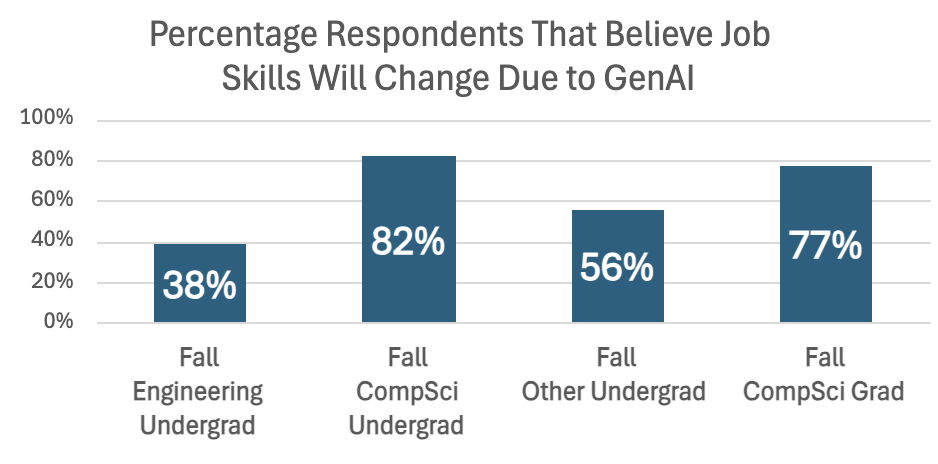}  % Replace with your figure file
    \vspace{-3pt}\caption{Responses on if job skills will change due to Gen AI.}
    \label{fig:skillchange1}
    \vspace{-7pt}
\end{figure}
 Of those that answered “Yes” to job skills changing, there was a follow up question of how soon they believe the changes will occur. Figure 4 shows computer science graduates had the highest expectation of changes occurring the soonest. Undergraduate Computer Science and Other Majors have a more balanced distribution, with nearly half predicting changes within 3 years and the rest expecting shifts within 5 years or longer. The Fall Undergraduate Engineering students mostly expect changes within 5 or 10 years and just 17\% expected changes to occur within 3 years.
\newline Figure 5 covers the remaining three questions on Gen AI preparedness. Engineering students reported the highest satisfaction with Gen AI discussions at 57\% and the greatest confidence that professors considered Gen AI at 64\%. In contrast, Computer Science students had the lowest satisfaction with 38\% and the least confidence in faculty engagement at 35\%. Despite this, confidence in curriculum preparation was high across all groups, with Engineering leading at 79\%. These results suggest stronger Gen AI integration in Engineering curricula, while Computer Science students perceive gaps in discussion and faculty consideration.
\begin{figure}[tbp!]
    \setlength{\belowcaptionskip}{-6pt} 
    \centering
 \includegraphics[width=0.45\textwidth]{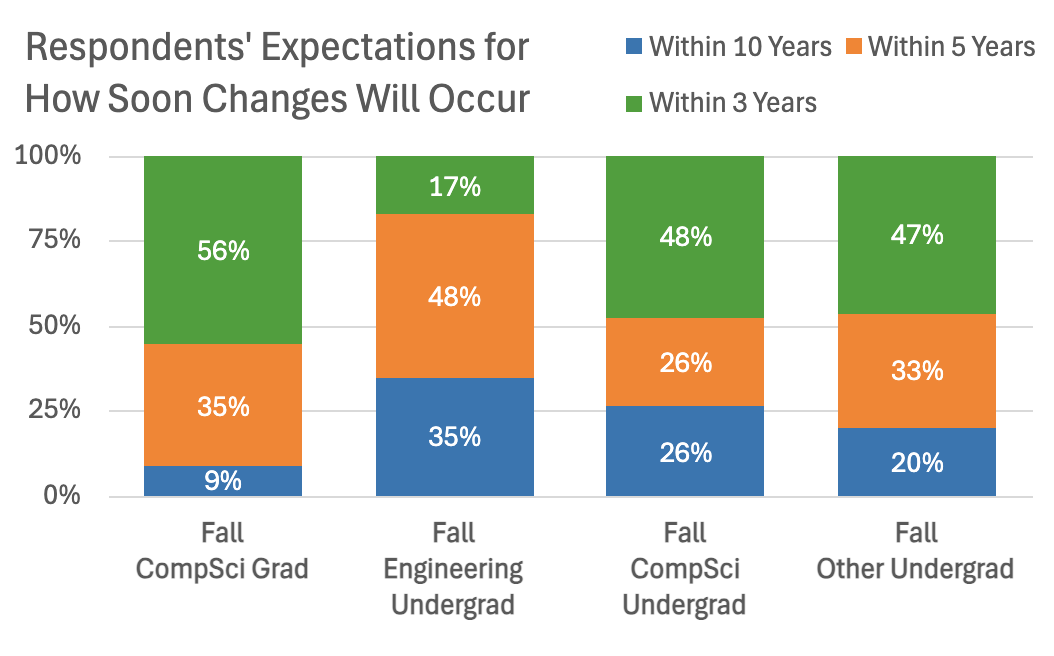}  % Replace with your figure file
 \vspace{-7pt}
    \caption{Respondent counts by degree and cohort.}
    \label{fig:skilltime}
\end{figure}
\begin{figure}[tbp!] 
    \centering
    \setlength{\belowcaptionskip}{-6pt} 
\includegraphics[width=0.48\textwidth]{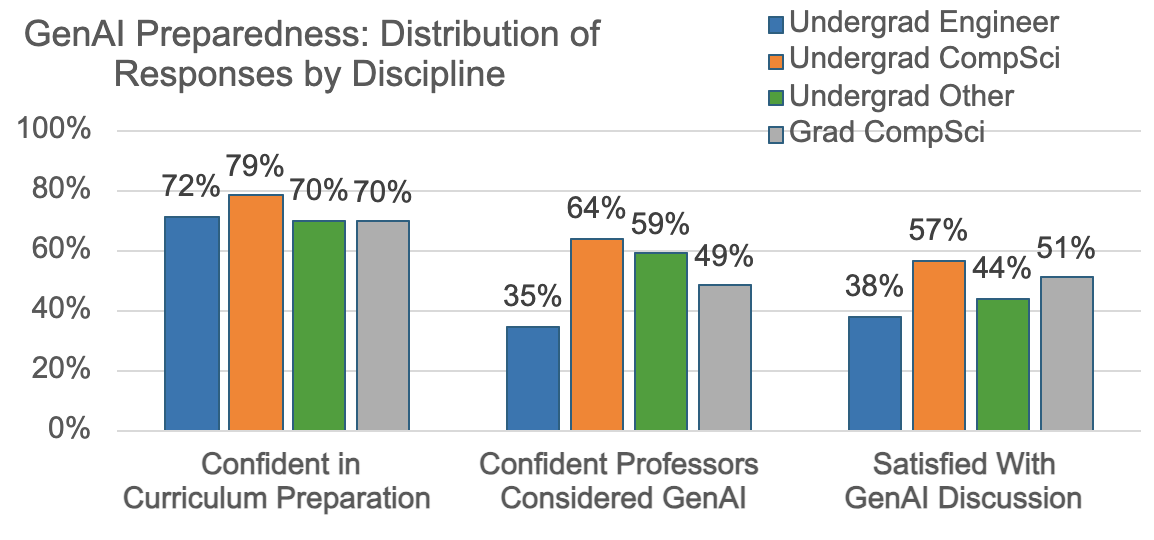}  % Replace with your figure file
\vspace{-7pt}
    \caption{Responses on preparedness}
    \label{fig:3prepare}
    \vspace{-8pt}
\end{figure}
\newline
\textbf{Concern}: 
The majority view Gen AI’s career impact positively, seen in Figure 6, with computer science graduates highest at 81\% and undergraduate Engineers, Computer Science, and Other lower with 70\%, 64\%, and 63\% respectively. Excitement about Gen AI benefits was high across all groups, peaking among Engineering at 77\%  and Computer Science Grads at 80\%. Other respondents notably lower at 67\%. Even with excitement and optimism about positive career impacts, concern was even greater across all  undergraduate degree categories. While less, the computer science graduate cohort still showed concern with 65\% concerned with Gen AI's. 
\begin{figure}[tbp!] 
    \centering
    \setlength{\belowcaptionskip}{-6pt} 
 \includegraphics[width=0.4\textwidth]{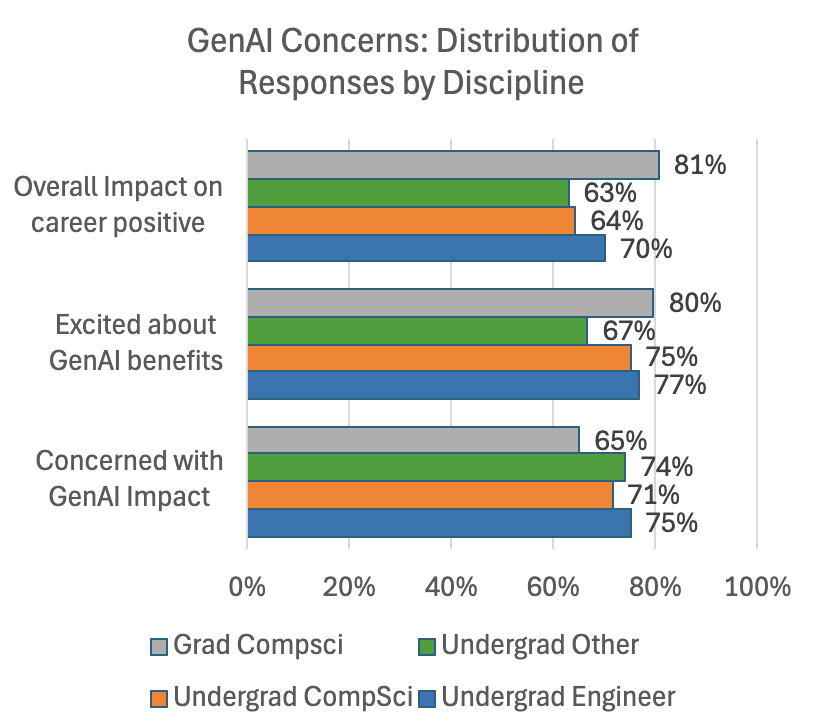}  % Replace with your figure file
 \vspace{-5pt}
    \caption{Responses on concern}
    \label{fig:concren}
    \vspace{-3pt}
\end{figure}
\newline
\textbf{Free Response}: 
In addition to the structured survey responses, a notable portion of participants contributed to the optional open-ended questions. These qualitative insights deepen understanding of the perspectives behind the patterns. 40–50\% of undergraduates and graduate students responded to the institutional recommendation prompt, and 23–25\% provided input on the additional thoughts prompt. 
\begin{figure}[tbp!]
    \centering
    \setlength{\belowcaptionskip}{-11pt} 
    \includegraphics[width=0.48\textwidth]{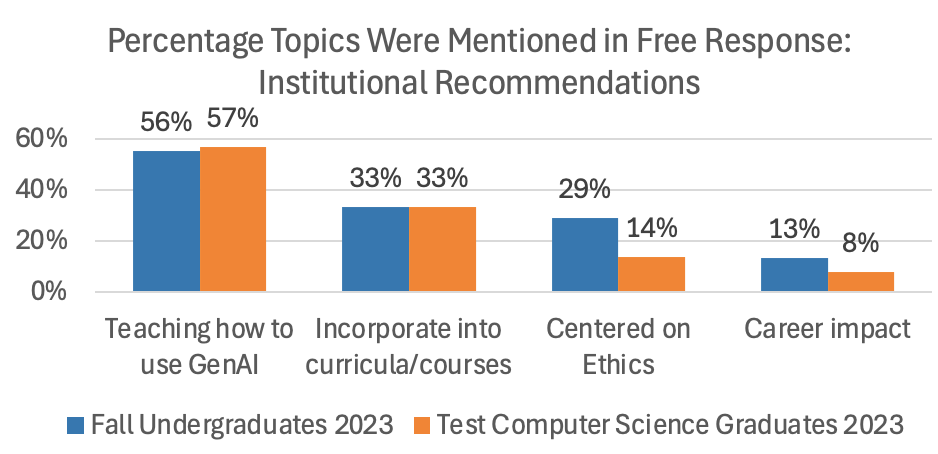}  % Replace with your figure file
    \caption{Topics raised in free response for institutional recommendations}
    \label{fig:free_topic_ir}
    \vspace{-6pt}
\end{figure}
The most common recommendation, mentioned by over 50\% for both cohorts, was teaching students how to use Gen AI tools (Figure 7). About 33\% of respondents also suggested integrating Gen AI into existing curricula or adding new courses. Ethics and career implications of Gen AI appeared more frequently in undergraduate responses, with ethics mentioned 15\% more and career impact 5\% more than in graduate responses.
\newline There were fewer who responded to the optional additional thoughts question, those who did showed clear emotional tone(Figure 8). Over half of undergraduate free responses expressed negative emotions, 29\% more than the graduate cohort, while graduates were 20\% more likely to specifically express positivity. Additional thoughts covered a broader range of themes(Figure 9). The computer science graduates frequently emphasized critical thinking, whereas undergraduates more often expressed uncertainty.
\begin{figure}[tbp!]
    \centering
    \includegraphics[width=0.48\textwidth]{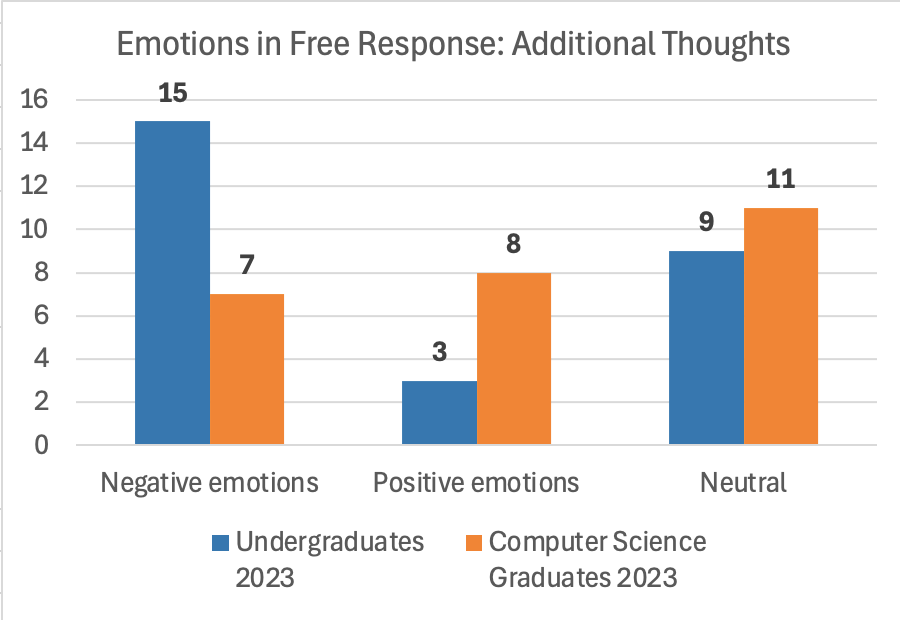}  % Replace with your figure file
    \caption{Percentage of free responses with either positive or negative emotions conveyed}
    \label{fig:emote}
\end{figure}
\begin{figure}[tb!]
    \centering
    \includegraphics[width=0.48\textwidth]{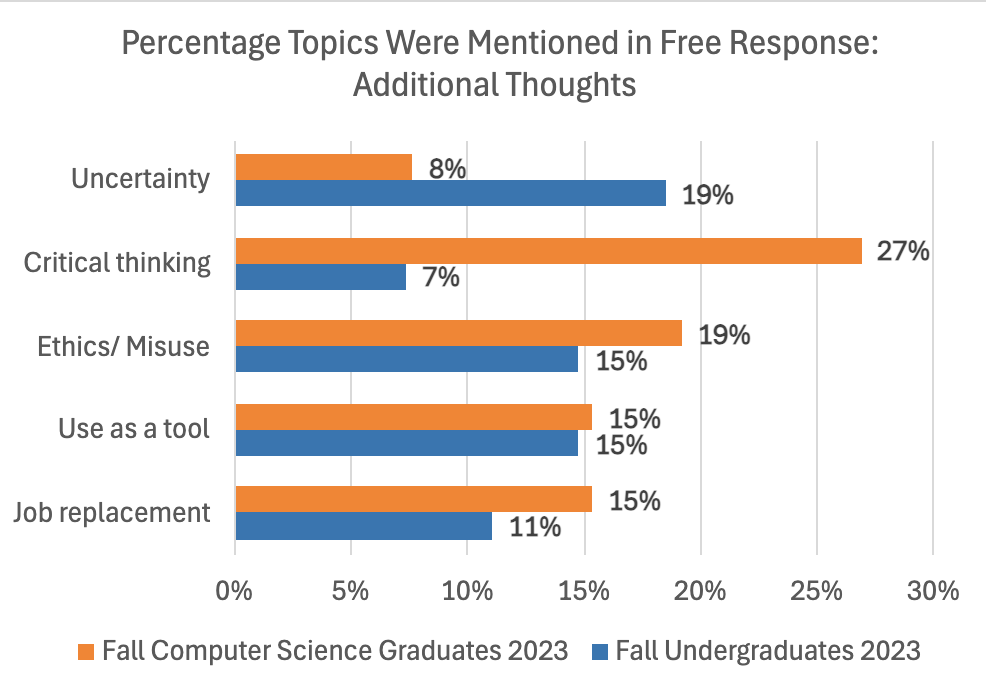}  % Replace with your figure file
    \vspace{-5pt}
    \caption{Topics raised in free response additional thoughts}
    \label{fig:addthot}
    \vspace{-15pt}
\end{figure}
These open-ended responses provide rich context to the survey data and are further explored in the Discussion. While the first prompt focused on Gen AI’s utility, the second invited any remaining reflections from participants.
\section{Discussion} 
The results have presented unique and shared perspectives across undergraduate programs, revealing areas of improvement. Although surveys were distributed uniformly, the potential for selection bias and low response frequencies remains a limitation. Response counts were modest compared to sample size, and some subgroups, particularly within non-STEM majors, were underrepresented. As a technology-focused institution, Georgia Tech’s student population may also have greater exposure to emerging technologies, including Gen AI, than students at other institutions. Even with these constraints, the findings offer important early insight into how Gen AI is shaping student experiences, perceptions, and expectations during a critical period of educational and technological transformation.
\newline\textbf{Awareness: the uncertainty in early adopters}
\newline While overall awareness was uniformly high, levels of understanding and confidence varied. Computer science undergraduates were the most likely to consider Gen AI a trend (21\%), whereas engineering and “Other” majors reported lower agreement (10–11\%). Despite this skepticism, nearly all respondents across disciplines viewed Gen AI as a meaningful advancement, with agreement ranging from 89\% to 100\%. Experience using Gen AI tools also varied, with graduate students reporting the highest exposure (93\%) and engineering undergraduates the lowest (78\%). Responses regarding the limitations of Gen AI’s potential were more common among engineering and graduate students. For the graduate cohort comprising of computer science, this could suggest increased exposure may come with greater critical awareness of its boundaries. For the undergraduate engineering students, skepticism may come from the belief that Gen AI's application is limited specifically for their degree.
These nuances are echoed in free responses, where students shared diverse sentiments:
\newline\textit{``Because aerospace engineering is pretty specialized and often times requires very difficult/complex work, I am ultimately unsure how Gen AI would be used in my field, if at all. However, I trust it can be a valuable tool in the future. I don't feel like Gen AI has any negative impacts on my future.''} — Undergraduate Engineer
\newline\vspace{-5pt}\newline\textit{``I think it has limited potential because Gen AI currently cannot distinct whether the information it gives is true or not.''} — Undergraduate Engineer
\newline\vspace{-5pt}\newline\textit{``It's very uncertain and that's what makes me slightly afraid. I also don't understand it very well, which I should work on.''} — Undergraduate Computer Science
\newline\vspace{-5pt}\newline As seen in some of these responses, there is skepticism of Gen AI achieving beyond what the current potential shows, lending to a potential experience of technology moving faster beyond students’ grasp. \textbf{Many students rely on self-directed learning}, coursework, and personal experiences to form their perspectives on Gen AI. This gap in structured AI education can lead to a divergence in awareness, confidence, and preparedness, with some students feeling optimistic and others expressing deep concern and uncertainty.
\newline In the optional free responses, 50\% of the students expressed negative emotions, including frustration, fear, and worry, while 20\% conveyed uncertainty, with overlapping sentiments highlighting the complex emotional landscape students navigate:
\newline\vspace{-5pt}\newline\textit{``I work in security. I am not excited about what's in store."} — Undergraduate Other
\newline\vspace{-5pt}\newline\textit{``I’m scared it’s going to make my career not what I want it anymore."} — Undergraduate Other
\newline\vspace{-5pt}\newline These comments reflect fear and uncertainty about Gen AI’s impact, amplified by a lack of classroom discussion. Without institutional guidance, students are left to navigate these changes on their own during critical educational years. Some expressed dissatisfaction with the current state of awareness in their programs overall:
\newline\vspace{-5pt}\newline\textit{``Even addressing its existence in class would be a step forward."} — Undergraduate Engineer
\newline\vspace{-5pt}\newline
These emotional responses and findings reveal a misalignment between perceived career impact and readiness and signal a broader \textbf{psychosocial landscape in flux} where students must construct academic identity and career expectations while navigating an unregulated and evolving technology. For a transformative tool like Gen AI, students are actively seeking guidance to understand its potential, limits, and ethical implications.
\newline
\textbf{Preparedness: addressing the elephant in the classroom}
\newline 
While students recognize Gen AI’s growing influence on their careers, many remain unsure whether their education is preparing them for the changes ahead. Unlike future students who may benefit from Gen AI-integrated curricula, the Gen AI Generation often relies on self-directed strategies to prepare for a Gen AI-driven workforce.
\newline  This preparedness gap is reflected in how different groups perceive Gen AI’s impact on job skills. There is a clear pattern that begins to emerge: those who feel prepared in their degree are most likely to expect a positive career impact. 
\newline  77\% of computer science graduates and 82\% computer science undergraduates anticipate job skill changes, only  56\% and 38\% of undergaduate other and engineering students anticipate such changes, with many engineering students expecting a slower pace, highlighting a lack of urgency in their field. These differences suggest the \textbf{need for tailored strategies that align with each group’s career context.}
\newline Despite these varied views, all cohorts reported relatively high confidence in their curricula, possibly reflecting trust in their degree or Institutions’ reputation. Still, engineering students stand out for their skepticism: they are the least likely to expect Gen AI-driven skill changes, report the lowest satisfaction with classroom discussions, and express the least confidence in faculty engagement. This gap signals a need for deeper integration and clearer guidance.
\newline The free response data adds depth to these findings. Despite overall confidence in their curriculum, over half of the free responses wanted to be taught \emph{how} to use Gen AI. Many also called for \textbf{stronger emphasis on fundamentals, critical thinking, and ethical considerations}:
\newline\vspace{-5pt}\newline\textit{``I think focusing on the idea of working alongside Gen AI on how to take advantage of it to produce better, faster results, but acknowledge its limitations} [...] \textit{ and discussing ethical and other ramifications of using AI in certain situations.} — Undergraduate Other
\newline\vspace{-5pt}\newline\textit{``It is a "tool" that can be used to aid students' work when used responsibly, and can be integrated in courses like other software tools, such as matlab, cad softwares, etc. but should not replace the fundamentals of engineering that is already being taught, such as hand calculations and understanding the theory."} — Undergraduate Engineer
\newline\vspace{-5pt}\newline\textit{``Strong implementation in to curriculum and learning tools to learn to work with AI and not treat it as a monster."} — Undergraduate Engineer
\newline\newline``[...]\textit{ the more adapted we are to it and prepared for its growth/advancement, the better off we'll be to use it as a tool, rather than as a toy.} — Undergraduate Other
\newline\vspace{-5pt}\newline
While many responses showed a proactive approach to Gen AI integration, others expressed skepticism or disappointment:
\newline\vspace{-5pt}\newline ``[...]\textit{I'm disappointed that tech won't prepare me for the near advent of those tools during my classes."} — Undergraduate Engineer
\newline\vspace{-5pt}\newline\textit{``As far as I can tell, it's a fad for almost all possible proposed use cases. Until stronger evidence comes to light for its all-encompassing power, it should be ignored in higher education."} — Undergraduate Engineer
\newline\vspace{-5pt}\newline\textit{``Show the downsides and flaws of things like gpt-4, discuss environmental impacts, and don’t use it. Gen AI is built on stolen data sets and harms artists. Stop endorsing its use.} — Undergraduate Other
\newline
\newline 
As Gen AI begins to enter technical curricula, students in other disciplines have an even greater risk being left behind. This gap underscores a fundamental pedagogical challenge: while students recognize Gen AI’s relevance, meaningful integration into their learning environments is falling behind, leaving students to independently navigate tools that carry both potential and risk. This lack of structure points to an urgent need for faculty guidance and training, intentional curriculum design, and interdisciplinary Gen AI literacy across academic programs. These efforts are essential to prepare a more equitable and capable generation for a Gen AI-driven future.  
\newline
\textbf{Concern: use at what cost?}
\newline 
Student concerns about Gen AI range from ethical implications and job security to institutional and faculty preparedness. While excitement is widespread, it is tempered by anxieties, reflecting the complexity of the Gen AI Generation’s experience. The 2023 Computer Science Graduate cohort reported the lowest concern (65\%) and the highest levels of excitement (80\%) and belief in Gen AI’s positive career impact (81\%). This likely reflects the integration of Gen AI into their curriculum and their confidence using such tools. 
\newline\vspace{-5pt}\newline``[...]\textit{ Gen AI may be able to do the bulk of the work when designing something functional and usable, so the focus can now be on designing technology that people like and that benefits them. I think Gen AI can help me do what I like, which would benefit people and society.``} — Graduate Computer Science
\newline\vspace{-5pt}\newline\textit{``I'm excited for the changes and technological advances it will bring.``} — Graduate Computer Science
\newline\vspace{-5pt}\newline In contrast, positive outlook drops to 64\% among undergraduate computer science students, \textbf{reinforcing how exposure shapes perception.}
\newline
Among undergraduates, belief in a positive career impact ranges from 63\% to 70\%, with engineering students on the higher end. Excitement is highest among computer science (75\%) and engineering students (77\%), while the “Other” cohort remains steady at 67\%. Even when students are unsure about Gen AI’s direct impact on their careers, many still express enthusiasm about its broader potential.
\newline\vspace{-5pt}\newline\textit{``I think less people should have to work for as long and more leisure and volunteering time will be helpful for mental health and balance.}" — Undergraduate Engineer
\newline\vspace{-5pt}\newline\textit{``Gen AI is the beginning of the 3rd “industrial” revolution. My life has changed a lot from ChatGPT.}" — Undergraduate Other
\newline\vspace{-5pt}\newline\textit{``I think it is important to emphasize both positives and negatives of Gen AI whenever mentioned. }" — Undergraduate Computer Science
\newline\vspace{-5pt}\newline Free responses reflect strong concern, especially in the open ended question additional thoughts prompt. \textbf{Over 50\% of undergraduates conveyed negative emotions}, with recurring themes of ethics, job replacement, and uncertainty, often expressed in long, passionate comments.
\newline\vspace{-5pt}\newline\textit{``I think it will actively damage the quality of work and data produced}[...]" — Undergraduate Engineer
\newline\vspace{-5pt}\newline\textit{``To be honest, I really do not care. I feel like Gen AI is going to obliterate most jobs. It almost seems inevitable.}[...]\textit{The only thing holding me from hopelessness is (blind) faith that my role as a researcher will not be rendered obsolete."} — Undergraduate Computer Science
\newline\vspace{-5pt}\newline\textit{``I think we are on a precipice of either ruin or greatness. Regardless of how we use it, Gen AI will become a powerful tool and have an impact greater than any singular tool since humans gained control of fire.}[...] "— Undergraduate Other
\newline\vspace{-5pt}\newline
Some students called for regulation:
\newline\vspace{-5pt}\newline``[...]\textit{we must ensure that there are precautions before the ai reaches a level where its misuse could pose a legitimate harm towards society and overall human advancement."} — Undergraduate Other
\newline\vspace{-5pt}\newline
Others expressed concern about government use: \newline\vspace{-5pt}\newline\textit{``I think the governments will start using Gen AI to produce propoganda in the coming years (if they haven't already) to manipulate public opinion. Or maybe just politicians with bot accounts blindly supporting them and arguing with dissenters on social media."} — Undergraduate Engineer
\newline\vspace{-5pt}\newline
And the more succinct, poignant response:
\newline\vspace{-5pt}\newline\textit{``Goodbye middle class, we doomed."} — Undergraduate Computer Science
\newline\vspace{-5pt}\newline
A compelling insight lies in who chose to share their perspectives in the optional free-response section. Among undergraduates, participation was highest among engineering and “other” majors, with response rates of 50\% and 44\%, respectively-compared to just 20\% among computer science undergraduate respondents. This could suggest that students who may not encounter Gen AI in their formal coursework still feel strongly about its impact and relevance.
\newline Student responses reveal strong concerns about the ethical, economic, and societal impacts of Gen AI, alongside a desire for greater guidance and critical engagement. Students are not only grappling with how to use Gen AI, but also with what kind of future it is shaping.
\section{Conclusion}
This study contributes to the growing discourse on generative artificial intelligence in education by examining how students are navigating both opportunity and uncertainty without consistent institutional guidance. It introduces the term Gen AI Generation to describe those in their formative educational years during the rise of Gen AI, students whose academic development is unfolding alongside the rapid adoption of these tools. Through analysis of over 250 survey responses, the findings reveal not only patterns of perceived readiness and ethical concern, but also a broader psychosocial landscape and pedagogical gap. Students are engaging with Gen AI amid unclear norms, forming academic identities and career expectations as institutions lag behind. These insights offer timely direction for educators and policymakers seeking to design inclusive, developmentally aligned, and ethically grounded approaches to Gen AI integration.
\newline
The Gen AI Generation faces a rapidly evolving landscape shaped by both excitement about Gen AI’s potential and concern about preparedness, ethics, and long-term implications. This study reveals a growing divide in student confidence and readiness, shaped more by exposure and self-directed integration of Gen AI than by formal curriculum. However, structured incorporation is inconsistent across disciplines, highlighting a broader institutional gap that must be addressed to ensure equitable support for all students navigating this technological shift. The divide is no longer simply between STEM and non-STEM fields, but between those with and without access to meaningful engagement with emerging technologies.
\newline
The Gen AI Generation is distinct in its developmental context. These students are not only adapting to powerful new tools, but doing so without established norms or guidance. As early adopters, they must balance traditional skills with new competencies while grappling with ethical dilemmas, shifting workforce expectations, and unknown cognitive impacts. Their adaptability is being tested in real time.
\newline
Students want more than access; they are asking for structure. They seek guidance on using Gen AI responsibly, developing critical thinking around AI-generated outputs, and building a strong foundation in ethical reasoning. \textbf{Their feedback calls for a holistic approach that prepares them not only to use Gen AI, but to question and shape it.}
\newline
As the first cohort to learn and grow in a Gen AI-centric world, the experiences of the Gen AI Generation will set the tone for future educational trajectories. Their voices point to an urgent need for Gen AI literacy that spans disciplines and prepares both students and faculty to engage with these tools thoughtfully and equitably. They envision a future where Gen AI is not a source of confusion or inequity, but a tool for growth, integrated into education in ways that are ethical, empowering, and human-centered.

\section{Future Considerations}
To better understand the evolving impact of Gen AI on student development and educational outcomes, future research should consider longitudinal studies that track student perceptions, readiness, and concerns over time. As Gen AI tools rapidly advance, students’ experiences and attitudes will likely shift in response to both technological innovation and institutional adaptation. Capturing these changes across semesters or academic years would provide valuable insight into how exposure, policy, and pedagogy shape preparedness and perception in the long term.
\newline In addition to tracking perception, researchers should examine the cognitive consequences of increased reliance on Gen AI tools for learning and problem-solving. Cognitive offloading may free students to focus on higher-order thinking, but it also raises questions about memory retention, deep learning, and critical reasoning. Understanding how sustained use of Gen AI affects cognitive engagement, information retention, and academic identity will be critical to developing pedagogical strategies that support, not replace, core learning processes.
\newline Expanding Gen AI literacy and integration across disciplines can help bridge the readiness gap and empower students across all domains to engage critically and confidently with this transformative technology.
\section*{Supplemental Materials}
Supplemental materials can be found under datasets  at https://huggingface.co/sirajmicaela
\section*{Acknowledgment}
Thank you to the  Georgia Tech Office of Academic Effectiveness for their support in providing the undergraduate cohort sample and sample methodology. And thank you to the participants who shared their voice.
\bibliographystyle{plain}  % or another style like apalike, IEEEtran, etc.
\bibliography{reference}   % no .bib extension here

\end{document}